\documentclass[prd,twocolumn,showpacs,showkeys,preprintnumbers,floatfix,
  nofootinbib]{revtex4-1}
\usepackage{amsfonts} 
\usepackage{amssymb} 
\usepackage{amsmath} 
\usepackage{subfigure} 
\usepackage{graphicx} 
\usepackage{array} 
\usepackage{dcolumn} 
\usepackage{bm} 
\usepackage{latexsym} 
\usepackage{longtable} 
\usepackage{hyperref} 
\usepackage{color}
\usepackage{mathrsfs}
\usepackage{comment}
\graphicspath{{./figs/}}

\newcommand{\CO}{{\cal O}}

\newcommand{\Tr}{\textrm{Tr}}

\newcommand{\mc}[1]{\mathcal{#1}}

\renewcommand{\mathcal}[1]{\mathscr{#1}}

\newif\ifContLineOne
\newif\ifContLineTwo
\newif\ifContLineThree

\def\conC#1{\vbox{\ialign{##\crcr
  \ifContLineThree\hrulefill\else\vphantom{\hrulefill}\fi\crcr
  \noalign{\kern3.2pt\nointerlineskip}
  \ifContLineTwo\hrulefill\else\vphantom{\hrulefill}\fi\crcr
  \noalign{\kern3.2pt\nointerlineskip}
  \ifContLineOne\hrulefill\else\vphantom{\hrulefill}\fi\crcr
  \noalign{\nointerlineskip}
  $\hfil\textstyle{\vbox to 14pt{}#1}\hfil$\crcr}}}

\def\DrawLeg#1#2{
  \kern-.2pt              
  \dimen2 =#1             
  \advance\dimen2 by 2pt  
  \dimen3 = 10.6pt        
  \dimen4 =3.6pt          
  \advance\dimen3 by -\dimen2 
  \multiply\dimen4 by #2
  \advance\dimen3 by \dimen4
  \raise\dimen2 \hbox{\vrule height\dimen3 width .4pt} 
  \kern-.2pt}             

\def\begC#1#2{\setbox0 =\hbox{$\textstyle{#2}$}
  \dimen0=.5\wd0 \dimen1=\ht0
  \conC{\hskip\dimen0}
  \count255=#1
  \ifnum\count255 =1 \ContLineOnetrue\else
  \ifnum\count255 =2 \ContLineTwotrue\else
  \ifnum\count255 =3 \ContLineThreetrue\fi\fi\fi
  \DrawLeg{\dimen1}{\count255}
  \conC{\hskip\dimen0}
  \kern-\dimen0\kern-\dimen0 \box0}

\def\endC#1#2{\setbox0 =\hbox{$\textstyle{#2}$}
  \dimen0=.5\wd0 \dimen1=\ht0
  \conC{\hskip\dimen0}
  \count255=#1
  \ifnum\count255 =1 \ContLineOnefalse\else
  \ifnum\count255 =2 \ContLineTwofalse\else
  \ifnum\count255 =3 \ContLineThreefalse\fi\fi\fi
  \DrawLeg{\dimen1}{\count255}
  \conC{\hskip\dimen0}
  \kern-\dimen0\kern-\dimen0 \box0}

\begin{document}

\title{
Taste non-Goldstone, flavor-charged pseudo-Goldstone boson decay constants in staggered chiral perturbation theory
}
\author{Jon A.~Bailey}
\author{Weonjong Lee}
\author{Boram Yoon}

\collaboration{SWME Collaboration}

\affiliation{
  Lattice Gauge Theory Research Center, FPRD, and CTP,\\
  Department of Physics and Astronomy,
  Seoul National University, Seoul, 151-747, South Korea
}
\date{\today}
\begin{abstract}
We calculate the axial current decay constants of taste non-Goldstone 
pions and kaons in staggered chiral perturbation theory through 
next-to-leading order.  
The results are a simple generalization of the results for the taste Goldstone case.  
New low-energy couplings are limited to analytic corrections that 
vanish in the continuum limit; certain coefficients of the chiral logarithms are modified, 
but they contain no new couplings.
We report results for quenched, fully dynamical, and partially quenched 
cases of interest in the chiral SU(3) and SU(2) theories.

\end{abstract}
\maketitle
%
\section{Introduction\label{sec:intr}}
The decay constants $f_\pi$ and $f_K$ parametrize hadronic matrix
elements entering the leptonic decays of $\pi$ and $K$ mesons.  The
values of the decay constants can be combined with the leptonic decay
rates from experiment to extract the CKM matrix elements $|V_{ud}|$
and $|V_{us}|$ and test first-row CKM unitarity.  Tighter constraints
on new physics are obtained by taking the ratio $f_K/f_\pi$ and the
form factor for the semileptonic decay $K\to\pi\ell\nu$ as theoretical
inputs; doing so has led to impressive agreement between the Standard
Model and experiment~\cite{Colangelo:2010et,Beringer:2012pdg}.


Staggered quarks have 4 tastes per flavor by construction 
\cite{Golterman:1984cy,Golterman:1985dz}.
The full taste symmetry group for a single massless flavor is $SU(4)_L \times SU(4)_R$
in the continuum limit ($a=0$).
At finite lattice spacing, lattice artifacts of $\mathcal{O}(a^2)$
break the taste symmetry, and the remaining exact chiral symmetry
is $U(1)_A$, which is enough to prevent the staggered quark mass 
from being additively renormalized.   
Hence, staggered fermions have an exact chiral symmetry at nonzero
lattice spacing.  In addition, lattice calculations with staggered fermions are
comparatively fast.  Staggered chiral perturbation theory (SChPT) was first
developed to describe the lattice artifacts and light quark mass
dependence of lattice data for the pseudo-Goldstone boson (PGB)
masses~\cite{Lee:1999zxa,Bernard:2001yj,Aubin:2003mg,Aubin:2003rg}.
Lattice data were extrapolated to the continuum limit and physical
quark masses to determine the light quark masses, tree-level PGB mass
splittings, and low-energy couplings (LECs); these served as inputs to
lattice calculations of the decay constants, semileptonic form
factors, mixing parameters, and other
quantities~\cite{Aubin:2004ck,Aubin:2004fs,Aubin:2004ej,Aubin:2005ar,Gray:2005ad,Aubin:2005zv,Okamoto:2005zg,Dalgic:2006dt,Aubin:2006xv,Bernard:2008dn,Bailey:2008wp,Gamiz:2009ku,Bazavov:2009bb,Bae:2010ki,Kim:2011qg,Bae:2011ff,Bazavov:2011aa,Bailey:2012rr,Bazavov:2012zs}.
Lattice calculations of $f_\pi$ have become precise enough to use it
to determine the lattice spacing~\cite{Bazavov:2009tw}.

While there have been a few attempts to calculate the decay constants for the
taste non-Goldstone sectors \cite{Aoki:1999av,Bernard:priv}, most lattice
calculations of the decay constants have been concentrated on the taste
Goldstone sector associated with the exact chiral symmetry of the staggered
action. 
%
%
%
In Ref.~\cite{Aubin:2003uc}, Aubin and Bernard calculated the decay constants
of the taste Goldstone pions and kaons through next-to-leading order (NLO) in
SChPT.  Here we extend their calculation to the
taste non-Goldstone pions and kaons; we find that the general results are
simply related to those in the taste Goldstone case.  We use the notation of
Ref.~\cite{SWME:2011aa} throughout.

In Sec.~\ref{sec:review} we recall the staggered chiral Lagrangian and the
tree-level propagators.  In Sec.~\ref{sec:decay_consts} we consider the
definition of the decay constants, recall the various contributions through NLO
in SChPT, and write down the general results in
the 4+4+4 theory.  Sec.~\ref{sec:results} contains the results for specific
cases of interest in the 1+1+1 theory, and we conclude in
Sec.~\ref{sec:conclusion}.  We use the notation of Ref.~\cite{SWME:2011aa}
throughout.

%
\section{\label{sec:review}Chiral Lagrangian for staggered quarks}
In this section, we write down the chiral Lagrangian for staggered quarks.
The single-flavor Lagrangian was formulated by Lee and Sharpe 
\cite{Lee:1999zxa} and generalized to multiple flavors by 
Aubin and Bernard \cite{Aubin:2003mg}.
Here, we consider the 4+4+4 theory, in which there are three flavors and 
four tastes per flavor. The exponential parameterization of the PGB fields
is a $12 \times 12$ unitary matrix,
\begin{align}
 \Sigma=e^{i\phi/f}\in \text{U(12)},
\end{align}
where the PGB fields are
\begin{align}
 \phi&=\sum_a \phi^a\otimes T^a,\\
 \phi^a&={
 \begin{pmatrix}
  U_a & \pi^+_a & K^+_a \\
  \pi^-_a & D_a & K^0_a \\
  K^-_a & \bar K^0_a & S_a
 \end{pmatrix}},\\
T^a&\in \{\xi_5,\ i\xi_{\mu5},\ i\xi_{\mu\nu}(\mu<\nu),\ \xi_\mu, \xi_I\}.
\label{eq:T^a}
\end{align}
Here $a$ runs over the 16 PGB tastes, and the $T^a$ are $4\times4$ generators
of $U(4)_T$; $\xi_I$ is the identity matrix.
Under a chiral transformation, $\Sigma$ transforms as
\begin{align}
 \text{SU}(12)_L\times \text{SU}(12)_R:\ \Sigma\rightarrow L\Sigma R^\dagger 
\end{align}
where $L,\ R\in \text{SU}(12)_{L,R}$.

In the standard power counting,
\begin{equation}
 \CO(p^2/\Lambda_\chi^2)
 \approx \CO(m_q/\Lambda_\chi)
 \approx \CO(a^2\Lambda_\chi^2)\,.
\label{eq:count}
\end{equation}
The order of a Lagrangian operator is defined as the sum of
$n_{p^2}$, $n_m$ and $n_{a^2}$, which are the number of derivative 
pairs, powers of (light) quark masses, and powers of the squared lattice spacing,
respectively, in the operator.
At leading order, the Lagrangian operators fall into three classes:
$(n_{p^2}, n_m, n_{a^2}) = (1,0,0)$, $(0,1,0)$ and $(0,0,1)$,
and we have
\begin{align}
 \label{F3LSLag}
 \mathcal{L}_\mathrm{LO} =
  &\frac{f^2}{8} \Tr(\partial_{\mu}\Sigma \partial_{\mu}\Sigma^{\dagger}) - 
    \frac{1}{4}\mu f^2 \Tr(M\Sigma+M\Sigma^{\dagger}) \nonumber\\
  &+ \frac{2m_0^2}{3}(U_I + D_I + S_I)^2 + a^2 (\mathcal{U+U^\prime})
\,,
\end{align}
where $f$ is the decay constant at leading order (LO), $\mu$ is the condensate parameter, and $M$ is the mass matrix,
\begin{equation}
M=
\begin{pmatrix}
m_u & 0 & 0 \\
0   & m_d & 0 \\
0 & 0 & m_s 
\end{pmatrix}
\otimes \xi_I.
\end{equation}
The term multiplied by $m_0^2$ is the anomaly contribution~\cite{Sharpe:2001fh},
and the potentials $\mathcal{U}$ and $\mathcal{U^\prime}$ are the taste symmetry breaking potentials of
Ref.~\cite{Aubin:2003mg}.

At NLO, there are six classes of operators satisfying $n_{p^2} + n_m + n_{a^2} = 2$,
but only two classes contribute to the decay 
constants: $(n_{p^2}, n_m, n_{a^2}) = $ $(1,1,0)$ and $(1,0,1)$.
Contributing operators in the former are Gasser-Leutwyler terms \cite{Gasser:1984gg},
\begin{align}
\label{eq:GLterms}
\mathcal{L}_\mathrm{GL}
 =& L_4\mathrm{Tr}(\partial_\mu\Sigma^\dagger\partial_\mu\Sigma)
       \mathrm{Tr}(\chi^\dagger\Sigma+\chi\Sigma^\dagger) \nonumber \\
  & \quad + L_5\mathrm{Tr}(\partial_\mu\Sigma^\dagger\partial_\mu\Sigma
       (\chi^\dagger\Sigma+\Sigma^\dagger\chi))
\,,
\end{align}
where $\chi = 2\mu M$, and contributing operators in the latter are $\mathcal{O}(p^2 a^2)$ 
terms enumerated by Sharpe and Van de Water \cite{Sharpe:2004is}.

\section{\label{sec:decay_consts}Decay constants of flavor-charged 
  pseudo-goldstone bosons}
For a flavor-charged PGB with taste $t$, $P^+_t$, the decay constant $f_{P_t^+}$ 
is defined by the matrix element of the axial current, $j_{\mu 5, t}^{P^+}$,
between the single-particle state and the vacuum:
\begin{equation}
\label{eq:def_decay_const}
 \langle 0 | j_{\mu 5, t}^{P^+} | P_t^+(p) \rangle  = -i f_{P_t^+} p_\mu.
\end{equation}
From the LO Lagrangian, the LO axial current is
\begin{equation}
 \label{eq:axial_curr}
 j_{\mu 5, t}^{P^+} 
  =  -i\frac{f^2}{8} \Tr 
   \left[ T^{t(3)} \mathcal{P}^{P^+} 
    (\partial_\mu \Sigma \Sigma^\dag + \Sigma^\dag \partial_\mu \Sigma)
   \right],
\end{equation}
where $T^{a(3)} \equiv I_3 \otimes T^a$, $I_3$is the identity matrix in flavor
space, and $\mathcal{P}^{P^+}$ is a projection operator that chooses the $P^+$
from the $\Sigma$ field.
For example, for $\pi^+$ it is $\mathcal{P}^{\pi^+}_{ij} = \delta_{i1}\delta_{j2}$.
In general, $\mathcal{P}^{P^+}_{ij} = \delta_{ix}\delta_{jy}$,
where $x$ and $y$ are the light quarks in $P^+$.  For flavor-charged states, $x\neq y$, by definition.
Note that $\mathcal{P}^{P^+}$ and $T^{a(3)}$ commute with each other.

\begin{figure}[t]
\centering
  \subfigure[]{
    \label{fig:waveftn_crxn}
    \includegraphics[width=8pc]{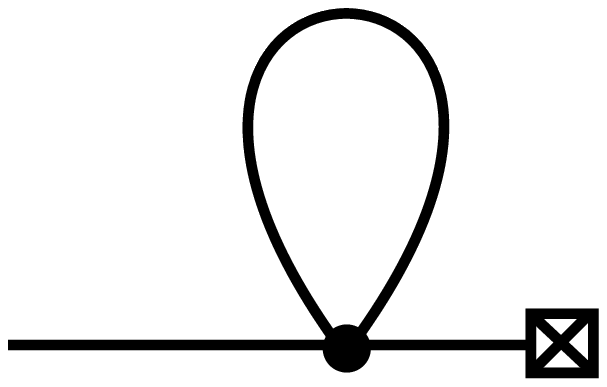}
  }
\qquad
  \subfigure[]{
    \label{fig:current_crxn}
    \includegraphics[width=6.2pc]{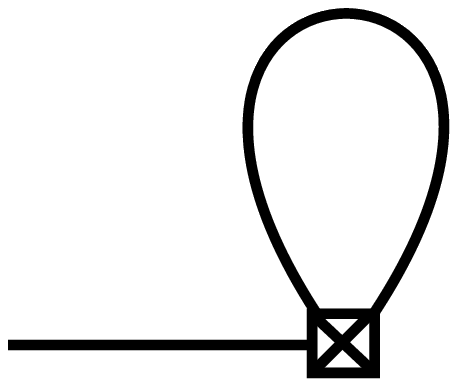}
  }
\caption{One-loop diagrams contributing to the decay constants at NLO.
  (a) is the wavefunction renormalization correction and
  (b) is the current correction.  The propagators include all insertions of hairpin vertices.}
\end{figure}

Expanding the exponentials $\Sigma=e^{i\phi/f}$ in the LO current gives 
\begin{align}
 &\partial_\mu \Sigma \Sigma^\dag + \Sigma^\dag \partial_\mu \Sigma
   = \frac{2i}{f} \partial_\mu \phi \nonumber \\
 & \qquad  - \frac{i}{3f^3}
   \left(
    \partial_\mu \phi \phi^2 - 2\phi \partial_\mu \phi \phi + \phi^2 \partial_\mu \phi
   \right)
  + \cdots.
\end{align}
The $\mathcal{O}(\phi)$ term of the axial current gives the LO term of
the decay constants, $f$, and NLO corrections from the wavefunction
renormalization. 
The wavefunction renormalization consists of NLO analytic terms
and one-loop chiral logarithms at NLO; we denote the former 
by $\delta f^{\textrm{anal},Z}_{P^+_t}$ and the latter
by $\delta f^Z_{P^+_t}$.
The $\mathcal{O}(\phi^3)$ term of the axial current also gives
one-loop chiral logarithms at NLO, $\delta
f^\textrm{current}_{P^+_t}$.  Figs.~\ref{fig:waveftn_crxn}
and~\ref{fig:current_crxn} show the one-loop corrections to the decay
constant.  In addition, there is an analytic contribution to the decay
constants from the NLO current. 
We denote the total of the NLO analytic terms by $\delta
f^\textrm{anal}_{P^+_t}$, which consists of $\delta
f^{\textrm{anal},Z}_{P^+_t}$ and analytic terms from the
NLO current.
Combining $\delta f^\textrm{anal}_{P^+_t}$ with the one-loop
corrections, we write the decay constants up to NLO:
\begin{equation}
 f_{P_t^+} = f\left[ 1 + \frac{1}{16\pi^2 f^2} 
  \left( 
    \delta f^Z_{P^+_t} + \delta f^\textrm{current}_{P^+_t} \right)
    + \delta f^\textrm{anal}_{P^+_t}
  \right]\label{eq:ftot}.
\end{equation}
In this section we outline the calculation of
$\delta f^Z_{P^+_t}$, $\delta f^\textrm{current}_{P^+_t}$,
and $\delta f^\textrm{anal}_{P^+_t}$ and present 
results for the 4+4+4 theory.

%
%
%

\subsection{\label{subsec:delta_f_z}Wavefunction renormalization correction}
At $\mathcal{O}(\phi)$ the axial current,
Eq.~\eqref{eq:axial_curr}, is
\begin{equation}
 j_{\mu 5, t}^{P^+, \phi} 
  = f \left( \partial_\mu \phi_{yx}^t \right),
\end{equation}
where we used $\tau_{ta} = 4\delta_{ta}$, 
$\mathcal{P}^{P^+}_{ij} = \delta_{ix}\delta_{jy}$,
and performed the trace over taste indices.
Here $\tau_{abcd\cdots}$ is defined by
\begin{equation}
 \tau_{abcd\cdots} \equiv \Tr( T^a T^b T^c T^d \cdots).
\end{equation}
The contributions of the $\mathcal{O}(\phi)$ current to the decay 
constants are defined by the matrix element
\begin{align}
 \langle 0 | j_{\mu 5, t}^{P^+, \phi} | P_t^+(p) \rangle
  &= f (-ip_\mu) \langle 0 | \phi_{yx}^t | P_t^+(p) \rangle \\
  &= f (-ip_\mu) \sqrt{Z_{P_t^+}},
\end{align}
where $Z_{P_t^+} = 1 + \delta Z_{P_t^+}$ is the wavefunction 
renormalization constant of the $\phi_{xy}^t$ field.
At NLO the wavefunction renormalization corrections are
\begin{equation}
 \frac{1}{16\pi^2 f^2}\delta f^Z_{P_t^+} + \delta f^{\textrm{anal},Z}_{P_t^+}
  =\frac{1}{2}\delta Z_{P_t^+}
  = - \frac{1}{2} \frac{d\Sigma}{dp^2} \bigg\vert_{p^2 = - m^2_{P_t^+}},
\end{equation}
where $\Sigma$ is the self-energy of $P_t^+$.
Using the self-energy from Ref.~\cite{SWME:2011aa}, we find the
one-loop corrections
\begin{align}
\label{eq:delta_f_z}
 \delta f
 &^Z_{P_t^+} 
  = \frac{1}{24} \sum_a 
   \Bigg[ 
    \sum_{Q} l(Q_a) \nonumber \\
 & + 16\pi^2 \int \frac{d^4 q}{(2\pi)^4} 
     \left(
      D_{xx}^a + D_{yy}^a - 2\theta^{at} D_{xy}^a
     \right)
   \Bigg],
\end{align}
where $Q$ runs over the six flavor combinations $xi$ and $yi$ for
$i\in\{u,d,s\}$, $a$ runs over the 16 PGB tastes in the $\mathbf{15}$
and $\mathbf{1}$ of $\text{SU}(4)_T$, $Q_a$ is the squared
tree-level pseudoscalar meson mass with flavor $Q$ and taste $a$,
and $\theta^{ab} \equiv \frac{1}{4}\tau_{abab}$.  In
Eq.~(\ref{eq:delta_f_z}), $l(Q_a)$ and $D^a_{ij}$ are chiral
logarithms and the disconnected piece of the tree-level propagator,
respectively~\cite{Aubin:2003mg,SWME:2011aa}:
\begin{align}
l(X)\equiv X\Bigl(\ln X/\Lambda^2 + \delta_1(\sqrt{X}L)\Bigr),
\end{align} 
where $\delta_1(\sqrt{X}L)$ is the finite volume correction of Ref.~\cite{Bernard:2001yj}, and
\begin{align}
D^a_{ij}&=-\frac{\delta_a}{(q^2+I_a)(q^2+J_a)}\nonumber \\
&\times\frac{(q^2+U_a)(q^2+D_a)(q^2+S_a)}{(q^2+\pi^0_a)(q^2+\eta_a)(q^2+\eta^\prime_a)}.\label{D_piece}
\end{align}
Here the names of mesons denote the squares of their tree-level masses,
and
\begin{align}
\delta_I = 4m_0^2/3,\quad&\delta_{\mu\nu} = 0,\quad\delta_5 = 0 \\
\delta_\mu = a^2\delta_V^\prime,&\quad\delta_{\mu5} = a^2\delta_A^\prime.\label{hpV_hpA}
\end{align}
For $X\in\{I,J,U,D,S\}$,
\begin{align}
X_a\equiv m_{X_a}^2 = 2\mu m_x + a^2\Delta_a,\label{eq:fntm}
\end{align}
where $m_x$ is the mass of the quark of flavor $x\in\{i,j,u,d,s\}$, while for
$X\in\{\pi^0,\eta,\eta^\prime\}$, the squares of the tree-level meson masses are the eigenvalues
of the matrix
\begin{equation}\label{444_mass_matrix}
        \begin{pmatrix}
        U_a +\delta_a & \delta_a & \delta_a \\
        \delta_a & D_a +\delta_a & \delta_a \\
        \delta_a & \delta_a & S_a +\delta_a
        \end{pmatrix}.
\end{equation}
The squared tree-level mass of a flavor-charged meson ($\pi^\pm,\ K^\pm, K^0, \bar{K}^0$) is
\begin{align}
P_t^+\equiv\frac{1}{2}(X_t+Y_t)=\mu(m_x+m_y)+a^2\Delta_t,\label{eq:tm}
\end{align}
where $X\ne Y\in\{U,D,S\}$ and $x\ne y\in\{u,d,s\}$.  The hairpin couplings
$\delta_{V,A}^\prime$ and taste splittings $\Delta_a$ are combinations of the
couplings of the LO Lagrangian~\cite{Aubin:2003mg}.

We defer discussing the analytic corrections $\delta f^{\textrm{anal},Z}_{P_t^+}$ 
to Sec.~\ref{subsec:analytic_cont}.

\subsection{\label{subsec:delta_f_current}Current correction}
The $\mathcal{O}(\phi^3)$ terms of the axial current are
\begin{align}
\label{eq:axial_curr_phi3}
 j_{\mu 5, t}^{P^+, \phi^3}
 =& - \frac{1}{24f} \tau_{tabc}  \Big(
  \partial_\mu \phi^a_{yk} \phi^b_{kl} \phi^c_{lx} \nonumber \\
 & \quad - 2 \phi^a_{yk} \partial_\mu \phi^b_{kl} \phi^c_{lx}
  + \phi^a_{yk} \phi^b_{kl} \partial_\mu \phi^c_{lx}
  \Big).
\end{align}
In the calculation of the matrix element defined in Eq.~\eqref{eq:def_decay_const}, 
each term of Eq.~\eqref{eq:axial_curr_phi3} contributes only one contraction
because the derivatively coupled fields in the current must contract with 
the external field to obtain a nonzero result.
For example, the first term gives 
\begin{equation}
 \partial_\mu \phi^a_{yk} \begC1{\phi^b_{kl}} \endC1{\phi^c_{lx}}
 \quad \rightarrow \quad -ip_\mu \delta^{ta}\delta^{bc} K_{xl,lx}^b,
\end{equation}
where~\cite{Sharpe:2000bc,Aubin:2003mg,SWME:2011aa}
\begin{align}
\label{Kintdef}
 K_{ij,kl}^a & \equiv \int \frac{d^4q}{(2\pi)^4}\langle\phi_{ij}^a\phi_{kl}^a\rangle
 \qquad \text{(no sum)},\\
\langle\phi^a_{ij}\phi^b_{kl}\rangle &=\delta^{ab}\left(\delta_{il}\delta_{jk}\frac{1}{q^2+\frac{1}{2}(I_a+J_a)}+\delta_{ij}\delta_{kl}D^a_{il}\right).\label{prop}
\end{align}

Collecting the three contributions from Eq.~\eqref{eq:axial_curr_phi3}, we find
\begin{align}
  i\frac{p_\mu}{6f} \sum_{a} 
  \left[
   \sum_{j} \left( K^a_{xj,jx} + K^a_{yj,jy} \right) 
   - 2\theta^{at} K^a_{xx,yy}
  \right],
\end{align}
where $j$ runs over $\{u,d,s\}$.
Performing the integrals over the loop momenta gives the one-loop current 
corrections to the decay constants:
\begin{align}
\label{eq:delta_f_curr}
 \delta f
 &_{P_t^+} ^{\textrm{current}}
  \equiv -\frac{1}{6} \sum_{a} 
  \Bigg[ 
   \sum_{Q} l(Q_a) \nonumber \\
 & + 16\pi^2 \int \frac{d^4q}{(2\pi)^4} 
    \left( D_{xx}^a + D_{yy}^a - 2\theta^{at} D_{xy}^a \right)
  \Bigg].
\end{align}
Note that $\delta f_{P_t^+}^{\textrm{current}}$ is proportional to 
the one-loop wavefunction renormalization correction, $\delta f_{P_t^+}^Z$.
This was shown in the taste Goldstone case in Ref.~\cite{Aubin:2003uc}.

%
%
\subsection{Next-to-leading order analytic contributions
            \label{subsec:analytic_cont}}
Now we consider the NLO analytic contributions to the decay constants.
They come from the $\mc{O}(p^2m)$ Gasser-Leutwyler Lagrangian in Eq.~(\ref{eq:GLterms})
and the $\mc{O}(p^2a^2)$ Sharpe-Van de Water Lagrangian of Ref.~\cite{Sharpe:2004is}.
Both Lagrangians contribute to wavefunction renormalization and the current.

The analytic corrections to the self-energy~\cite{SWME:2011aa}
give the wavefunction renormalization correction
\begin{align}
 \delta f_{P_t^+}^{\textrm{anal},Z}
  =& - \frac{64}{f^2} L_4 \mu (m_u + m_d + m_s)  \nonumber \\
  & - \frac{8}{f^2} L_5 \mu (m_x + m_y)-\frac{8}{f^2} a^2 \mc{C}_t,
\end{align}
while the NLO current from the Gasser-Leutwyler
terms gives the current correction
\begin{align}
 \delta f_{P_t^+}^{\textrm{current}, \textrm{GL}} 
  =& \frac{128}{f^2} L_4 \mu (m_u + m_d + m_s)  \nonumber \\
   & \qquad + \frac{16}{f^2} L_5 \mu (m_x + m_y).
\end{align}

The contributions of the $\mathcal{O}(p^2 a^2)$ operators coming from
the Sharpe-Van de Water Lagrangian in Ref.~\cite{Sharpe:2004is} give
the current correction
\begin{align}
 \delta f_{P_t^+}^{\textrm{current}, \textrm{SV}} = a^2\mc{C}^\prime_t.
\end{align}
The LECs $\mc{C}_t$ and $\mc{C}^\prime_t$ are degenerate within 
the irreps of the lattice symmetry group.
Sharpe and Van de Water observed that contributions from the $\mathcal{O}(p^2 a^2)$
source operators destroy would-be relations 
between the SO(4)-violations in the PGB masses and the (axial current)
decay constants~\cite{Sharpe:2004is}.

Collecting the analytic corrections, we have (in the 4+4+4 theory)
\begin{align}
\label{eq:delta_f_anal}
 \delta f_{P_t^+}^{\textrm{anal}} 
  =& \frac{64}{f^2} L_4 \mu (m_u + m_d + m_s)  \nonumber \\
   & \qquad + \frac{8}{f^2} L_5 \mu (m_x + m_y) 
      + a^2 \mathcal{F}_t,
\end{align}
where the constants $\mc{F}_t$ subsume the constants $\mc{C}^{(\prime)}_t$.  Examining Eqs.~(\ref{eq:delta_f_z}), (\ref{eq:delta_f_curr}), and (\ref{eq:delta_f_anal}), we see that the constants $\mc{F}_t$ (for $t\neq 5$) are the only new LECs entering the (NLO) expressions for the decay constants, in the sense that the others are present also in the taste Goldstone case.

\section{\label{sec:results}Results}
To formulate the full QCD and (partially) quenched results in
rooted SChPT, we employ the replica 
method \cite{Bernard:1993sv, Damgaard:2000gh, Bernard:2007ma}.
Rooting introduces a factor of $1/4$ in front of the explicit chiral logarithms $l(Q_a)$ in
Eqs.~\eqref{eq:delta_f_z} and \eqref{eq:delta_f_curr} and in the $L_4$ term 
in Eq.~\eqref{eq:delta_f_anal}.  We must also replace the eigenvalues of the mass matrix~\eqref{444_mass_matrix} with those of the matrix obtained by sending $\delta_a\to\delta_a/4$ there.  We have
\begin{align}
\label{eq:delta_f_rooted}
 \delta f_{P_F^+}
  &= \delta f_{P_F^+}^{Z} + \delta f_{P_F^+}^{\textrm{current}}=\delta f_{P_F^+}^{\textrm{con}} + \delta f_{P_F^+}^{\textrm{disc}}, \\
 \label{eq:delta_f_anal_rooted}
 \delta f_{P_t^+}^{\textrm{anal}} 
  &= \frac{16}{f^2} L_4 \mu (m_u + m_d + m_s)  \nonumber \\
  &  \qquad + \frac{8}{f^2} L_5 \mu (m_x + m_y) 
      + a^2 \mathcal{F}_t,
\end{align}
where
\begin{align}
 \label{eq:delta_f_con}
 \delta f_{P_F^+}^{\textrm{con}}
  &\equiv -\frac{1}{32} \sum_{Q,B} g_B ~ l(Q_B), \\
 \label{eq:delta_f_disc}
 \delta f_{P_F^+}^{\textrm{disc}}
  &\equiv -2\pi^2 \int \frac{d^4q}{(2\pi)^4} 
    \Big( 
     D_{xx}^I + D_{yy}^I - 2 D_{xy}^I \nonumber \\
  & \qquad \qquad + 4 D_{xx}^V + 4 D_{yy}^V - 2\Theta^{VF} D_{xy}^V 
   \nonumber \\
  & \qquad \qquad  + 4 D_{xx}^A + 4 D_{yy}^A - 2\Theta^{AF} D_{xy}^A
    \Big)
\,.
\end{align}
In Eq.~\eqref{eq:delta_f_disc}, the flavor-neutral, tree-level masses ($\pi^0_a,\ \eta_a,\ \eta^\prime_a$) appearing in $D^a_{ij}$ have been replaced with the masses obtained by sending $\delta_a\to\delta_a/4$ in the flavor-neutral meson mass matrix.
In Eqs.~(\ref{eq:delta_f_con}) and (\ref{eq:delta_f_disc}), we summed over $a$ within each SO(4) irrep in Eqs.~\eqref{eq:delta_f_z} and
\eqref{eq:delta_f_curr}, 
$B$ and $F$ represent (taste) SO(4) irreps,
\begin{equation}
 B, F \in \{ I, V, T, A, P \},
\end{equation}
$t \in F$, and
\begin{equation}
 \Theta^{BF} \equiv \sum_{a \in B} \theta^{at}, \quad
 g_B \equiv \sum_{a \in B} 1.\label{eq:coeff}
\end{equation}
The coefficients $\Theta^{BF}$ are given in Table~\ref{tab:coeff}.
\begin{table*}[htbp]
\caption{\label{tab:coeff}The coefficient $\Theta^{BF}$ defined
  in Eq.~(\ref{eq:coeff}) is in row $B$ and column $F$.}
\begin{center}
\begin{tabular}{c|rrrrr}
\hline\hline
$B \backslash F$ & $P$ & $A$ & $T$ & $V$ & $I$ \\
\hline
$V$ & $-4$ & $2$ & $0$ & $-2$ & $\phantom{-}4$ \\
$A$ & $-4$ & $-2$ & $0$ & $2$ & $4$ \\
\hline\hline
\end{tabular}
\end{center}
\end{table*}
The loop corrections differ from those in the taste
Goldstone case only in the values of the coefficients $\Theta^{BF}$.

Eq.~\eqref{eq:delta_f_anal_rooted} subsumes the NLO analytic corrections in fully dynamical and partially quenched SU(3) SChPT; in the former case, $m_x\neq m_y$ are chosen from $m_u$, $m_d$, and $m_s$.  In the quenched case, the $L_4$ term is dropped.  To obtain the NLO analytic corrections in SU(2) SChPT, we drop terms with the heavy quark mass(es), and the LECs become heavy quark mass dependent~\cite{Bailey:2012wb}.

Below we give the one-loop contributions to the decay 
constants for each of these cases.
In Sec.~\ref{subsubsec:su3_full} and Sec.~\ref{subsubsec:su3_pq},
fully dynamical and partially quenched results for the 1+1+1 and 
2+1 flavor cases in SU(3) chiral perturbation theory are given.
The analogous results in SU(2) chiral perturbation theory
are presented in Sec.~\ref{subsec:su2}.
In Sec.~\ref{subsubsec:su3_q}, we write down the results in the quenched case.

\subsection{\label{subsec:su3}SU(3) chiral perturbation theory}
%
\subsubsection{\label{subsubsec:su3_full}Fully dynamical case}
In Eq.~\eqref{eq:delta_f_con} $Q$ runs over six flavor combinations,
$xi$ and $yi$ for $i \in \{u, d, s\}$. 
Setting $xy = ud,\ us,\ ds$ gives the results for the $\pi^+$, $K^+$, and $K^0$ 
in the fully dynamical 1+1+1 flavor case:
\begin{align}
 \delta f_{\pi_F^+}^{\textrm{con}} 
  &= -\frac{1}{32} \sum_{B} g_B \Big( l(U_B) + 2l(\pi_B^+) \nonumber \\
  & \qquad \qquad    + l(K_B^+) + l(D_B) + l(K_B^0) \Big), \label{eq:f_pi_con_su3_full}
\end{align}
\begin{align}
 \delta f_{K_F^+}^{\textrm{con}} 
  &= -\frac{1}{32} \sum_{B} g_B \Big( l(U_B) + l(\pi_B^+) \nonumber \\
  & \qquad \qquad    + 2l(K_B^+) + l(K_B^0) + l(S_B) \Big). \label{eq:f_K_con_su3_full}
\end{align}
\begin{align}
 \delta f_{K_F^0}^{\textrm{con}} 
  &= -\frac{1}{32} \sum_{B} g_B \Big( l(\pi_B^+) + l(D_B) \nonumber \\
  & \qquad \qquad    + 2l(K_B^0) + l(K_B^+) + l(S_B) \Big). \label{eq:f_Kn_con_su3_full}
\end{align}
In the disconnected parts, Eq.~\eqref{eq:delta_f_disc}, the integrals 
can be performed as explained in Ref.~\cite{Aubin:2003mg}.
After performing the integrals and decoupling the $\eta_I'$ by taking 
$m_0^2 \rightarrow \infty$~\cite{Sharpe:2001fh}, 
we find
\begin{align}
 \delta f
 &_{\pi_F^+}^{\textrm{disc}}
  = \sum_{X} \Bigg[
   \frac{1}{6} \Big\{ 
    R^{DS}_{U \pi^0 \eta} (X_I) l(X_I) \nonumber \\
 & \quad + R^{US}_{D \pi^0 \eta} (X_I) l(X_I) 
         - 2 R^S_{\pi^0 \eta} (X_I) l(X_I)
   \Big\} \nonumber \\
 & + \frac{1}{4} a^2 \delta_V' \Big\{ 
    2R^{DS}_{U \pi^0 \eta \eta'} (X_V) l(X_V) \nonumber \\
 & \quad + 2R^{US}_{D \pi^0 \eta \eta'} (X_V) l(X_V) 
         - \Theta^{VF} R^S_{\pi^0 \eta \eta'} (X_V) l(X_V)
   \Big\} \nonumber \\
 & + (V \rightarrow A) 
  \Bigg],
\label{eq:f_pi_disc_su3_full}
\end{align}
\begin{align}
 \delta f
 &_{K_F^+}^{\textrm{disc}}
  = \sum_{X} \Bigg[
   \frac{1}{6} \Big\{ 
    R^{DS}_{U \pi^0 \eta} (X_I) l(X_I) \nonumber \\
 & \quad + R^{UD}_{S \pi^0 \eta} (X_I) l(X_I) 
         - 2 R^D_{\pi^0 \eta} (X_I) l(X_I)
   \Big\} \nonumber \\
 & + \frac{1}{4} a^2 \delta_V' \Big\{ 
    2R^{DS}_{U \pi^0 \eta \eta'} (X_V) l(X_V) \nonumber \\
 & \quad + 2R^{UD}_{S \pi^0 \eta \eta'} (X_V) l(X_V) 
         - \Theta^{VF} R^D_{\pi^0 \eta \eta'} (X_V) l(X_V)
   \Big\} \nonumber \\
 & + (V \rightarrow A) 
  \Bigg],
\label{eq:f_K_disc_su3_full}
\end{align}
\begin{align}
 \delta f
 &_{K_F^0}^{\textrm{disc}}
  = \sum_{X} \Bigg[
   \frac{1}{6} \Big\{ 
    R^{US}_{D \pi^0 \eta} (X_I) l(X_I) \nonumber \\
 & \quad + R^{UD}_{S \pi^0 \eta} (X_I) l(X_I) 
         - 2 R^U_{\pi^0 \eta} (X_I) l(X_I)
   \Big\} \nonumber \\
 & + \frac{1}{4} a^2 \delta_V' \Big\{ 
    2R^{US}_{D \pi^0 \eta \eta'} (X_V) l(X_V) \nonumber \\
 & \quad + 2R^{UD}_{S \pi^0 \eta \eta'} (X_V) l(X_V) 
         - \Theta^{VF} R^U_{\pi^0 \eta \eta'} (X_V) l(X_V)
   \Big\} \nonumber \\
 & + (V \rightarrow A) 
  \Bigg].
\label{eq:f_Kn_disc_su3_full}
\end{align}
In the sum, $X$ runs over the subscripts of the residue, $R$,
where the residues are defined by
\begin{align}
 R_{B_1B_2\cdots B_n}^{A_1A_2\cdots A_k}(X_F)
  \equiv\frac{\prod_{A_j}(A_{jF}-X_F)}{\prod_{B_{i}\ne X}(B_{iF}-X_F)},
\end{align}
where $X\in\{B_1,B_2,\dots,B_n\}$ and $F\in\{V,A,I\}$ is the $SO(4)_T$ irrep.

The results in the $2+1$ flavor case are easily obtained by
setting $xy = ud,\ us$ and $m_u = m_d$.
Eq.~\eqref{eq:delta_f_con} gives connected contributions
for the $\pi$ and $K$:
\begin{align}
 \label{eq:f_pi_con_su3_full_21}
 \delta f_{\pi_F}^{\textrm{con}} 
  &= -\frac{1}{16} \sum_{B} g_B \Big( 2l(\pi_B) + l(K_B) \Big), \\
 \label{eq:f_K_con_su3_full_21}
 \delta f_{K_F}^{\textrm{con}} 
  &= -\frac{1}{32} \sum_{B} g_B \Big( 2l(\pi_B) + 3l(K_B) + l(S_B) \Big).
\end{align}
Setting $xy = ud,\ us$ and $m_u = m_d$ in Eq.~\eqref{eq:delta_f_disc} gives
\begin{align}
 \label{eq:delta_f_pi_su3_21_a}
 \delta f
  &_{\pi_F}^{\textrm{disc}} = 
  \frac{1}{4} a^2 \delta_V'(4-\Theta^{VF}) \sum_X R^S_{\pi \eta \eta'} (X_V) l(X_V)
    \nonumber \\
  &\qquad\qquad + (V \rightarrow A),
\end{align}
and
\begin{align}
 \label{eq:delta_K_pi_su3_21_a}
 \delta f
  &_{K_F}^{\textrm{disc}} = 
    \frac{1}{6}\sum_X \Big\{ R^S_{\pi \eta} (X_I) l(X_I) + R^\pi_{S \eta} (X_I) l(X_I)\Big\}
    \nonumber \\
  & \qquad \qquad \qquad -2 l(\eta_I)
    \nonumber\\ 
  & +\frac{1}{4} a^2 \delta_V' \sum_X \Big\{ 
    2R^S_{\pi \eta \eta'} (X_V) l(X_V) + 2R^\pi_{S \eta \eta'} (X_V) l(X_V)
    \nonumber \\
  & \qquad \qquad \qquad - \Theta^{VF} R_{\eta \eta'}(X_V) l(X_V)  \Big \}
    \nonumber \\
  & + (V \rightarrow A) \,,
\end{align}
where $R_{B_1 B_2}(X_F)$ is defined by
\begin{align}
 R_{B_1 B_2}(X_F)
  &= 
    \begin{cases} 
     \dfrac{1}{B_2 - B_1} \quad (X_F = B_1) \\
     \dfrac{1}{B_1 - B_2} \quad (X_F = B_2)
    \end{cases}.
\end{align}
Using the tree-level masses of the taste singlet channel, one finds
\begin{align}
 R_{\pi\eta}^{S}(\pi_I)=\frac{3}{2},& \quad
  R_{\pi\eta}^S(\eta_I)=-\frac{1}{2},\\
 R_{S\eta}^\pi(S_I)=3,&\quad
  R_{S\eta}^\pi(\eta_I)=-2.
\end{align}
They simplify the results, Eqs.~\eqref{eq:delta_f_pi_su3_21_a} 
and \eqref{eq:delta_K_pi_su3_21_a}:
\begin{align}
 \label{eq:delta_f_pi_su3_21_b}
 \delta f
  &_{\pi_F}^{\textrm{disc}} = 
  \frac{1}{4} a^2 \delta_V' (4-\Theta^{VF}) \Bigg[ 
    \frac{S_V - \pi_V}{(\eta_V - \pi_V)(\eta_V' - \pi_V)} l(\pi_V)
    \nonumber \\
  & \qquad + \frac{S_V - \eta_V}{(\pi_V - \eta_V)(\eta_V' - \eta_V)} l(\eta_V)
    \nonumber \\
  & \qquad + \frac{S_V - \eta_V'}{(\pi_V - \eta_V')(\eta_V - \eta_V')} l(\eta_V')
    \Bigg]
    \nonumber \\
  & + (V \rightarrow A),
\end{align}
\begin{align}
 \label{eq:delta_K_pi_su3_21_b}
 \delta f
  &_{K_F}^{\textrm{disc}} = 
    \frac{1}{12} \Big[ 3 l(\pi_I) - 5 l(\eta_I) + 6 l(S_I) 
      - 4 l(\eta_I) \Big]
    \nonumber \\
  & + \frac{1}{2} a^2 \delta_V' \Bigg[ 
    \frac{S_V - \pi_V}{(\eta_V - \pi_V)(\eta_V' - \pi_V)} l(\pi_V)
    \nonumber \\
  & \qquad + \frac{(\pi_V - \eta_V)^2 + (S_V - \eta_V)^2}
      {(\pi_V - \eta_V)(\eta_V' - \eta_V)(S_V - \eta_V)} l(\eta_V)
    \nonumber \\
  & \qquad + \frac{(\pi_V - \eta_V')^2 + (S_V - \eta_V')^2}
      {(\pi_V - \eta_V')(\eta_V - \eta_V')(S_V - \eta_V')} l(\eta_V')
    \nonumber \\
  & \qquad + \frac{\pi_V - S_V}{(\eta_V - S_V)(\eta_V' - S_V)} l(S_V)
    \nonumber \\
  & \qquad - \frac{1}{2} \Theta^{VF} \frac{1}{\eta_V - \eta_V'}
      \Big\{ l(\eta_V') - l(\eta_V) \Big\}
    \Bigg]
    \nonumber \\
  & + (V \rightarrow A).
\end{align}
%

\subsubsection{\label{subsubsec:su3_pq}Partially quenched case}
In the partially quenched case, the valence quark masses, $m_x$ and $m_y$
are not degenerate with the sea quark masses, $m_u$, $m_d$ and $m_s$.
The connected contributions to the decay constants in the partially quenched 
1+1+1 flavor case are 
\begin{align}
 \label{eq:delta_f_con_su3_pq}
 \delta f_{P_F^+}^{\textrm{con}}
  = -\frac{1}{32} \sum_{Q,B} g_B ~ l(Q_B).
\end{align}

Performing the integrals in Eq.~\eqref{eq:delta_f_disc} keeping all
quark masses distinct gives the disconnected contributions for the 
partially quenched 1+1+1 flavor case:
\begin{align}
 \label{eq:delta_f_disc_su3_pq_a}
 \delta f
 &_{P_F^+, m_x \neq m_y}^{\mathrm{disc}} = \sum_{Z} \Bigg[
  \frac{1}{6} \Big\{ 
  D^{UDS}_{X \pi^0 \eta, X}(Z_I) l(Z_I)
  \nonumber \\
 &\qquad + D^{UDS}_{Y \pi^0 \eta, Y}(Z_I) l(Z_I)
  - 2 R^{UDS}_{XY \pi^0 \eta}(Z_I) l(Z_I) 
  \Big\} \nonumber \\
 &\quad + \frac{1}{4} a^2 \delta_V' \Big\{ 
  2D^{UDS}_{X \pi^0 \eta \eta', X}(Z_V) l(Z_V)
  \nonumber \\
 &\qquad + 2D^{UDS}_{Y \pi^0 \eta \eta', Y}(Z_V) l(Z_V)
  \nonumber \\
 &\qquad - \Theta^{VF} R^{UDS}_{XY \pi^0 \eta \eta'}(Z_V) l(Z_V) 
  \Big\} 
  + (V \rightarrow A)
 \Bigg] \nonumber \\
 & + \frac{1}{6} \Big\{ 
    R^{UDS}_{X \pi^0 \eta}(X_I) \tilde{l}(X_I) 
  + R^{UDS}_{Y \pi^0 \eta}(Y_I) \tilde{l}(Y_I)
  \Big\} \nonumber \\
 & + \frac{1}{2} a^2 \delta_V' \Big\{
    R^{UDS}_{X \pi^0 \eta \eta'}(X_V) \tilde{l}(X_V)
  + R^{UDS}_{Y \pi^0 \eta \eta'}(Y_V) \tilde{l}(Y_V)
  \Big\} \nonumber \\
 & + (V \rightarrow A),
\end{align}
where
\begin{align}
D_{B_1B_2\cdots B_n,B_i}^{A_1A_2\cdots A_k}(X_F)
\equiv -\frac{\partial}{\partial B_{iF}}
R_{B_1B_2\cdots B_n}^{A_1A_2\cdots A_k}(X_F)
\end{align}
and
\begin{align}
\tilde{l}(X)\equiv -\Bigl(\ln X/\Lambda^2 + 1\Bigr) + \delta_3(\sqrt{X}L).
\end{align}
Here $\delta_3(\sqrt{X}L)$ is the finite volume correction defined in
Ref.~\cite{Bernard:2001yj},
and $X$ and $Y$ represent the squared tree-level masses of $x\bar{x}$
and $y\bar{y}$ PGBs, respectively.

For $m_x = m_y$, we find
\begin{align}
 \label{eq:delta_f_disc_su3_pq_b}
 \delta f
 &_{P_F^+, m_x = m_y}^{\mathrm{disc}} = 
 \frac{1}{4} a^2 \delta_V' (4-\Theta^{VF}) \Bigg[
   R^{UDS}_{X \pi^0 \eta \eta'}(X_V) \tilde{l}(X_V)
  \nonumber \\
 & \qquad \qquad \qquad \qquad 
  + \sum_Z D^{UDS}_{X \pi^0 \eta \eta', X}(Z_V) l(Z_V)
 \Bigg] \nonumber \\
 &+(V \rightarrow A).
\end{align}

The connected contributions in the 2+1 flavor case are obtained by setting 
$m_u = m_d$ in Eq.~\eqref{eq:delta_f_con_su3_pq}.
To obtain the disconnected contributions, we perform the integrals
in Eq.~\eqref{eq:delta_f_disc} setting $m_u = m_d$.
For $m_x \neq m_y$, we find
\begin{align}
 \label{eq:delta_f_xney_pq_21}
 \delta f
 &_{P_F^+, m_x \neq m_y}^{\mathrm{disc}} = \sum_{Z} \Bigg[
  \frac{1}{6} \Big\{ 
  D^{\pi S}_{X \eta, X}(Z_I) l(Z_I)
  \nonumber \\
 &\qquad + D^{\pi S}_{Y \eta, Y}(Z_I) l(Z_I)
  - 2 R^{\pi S}_{XY \eta}(Z_I) l(Z_I) 
  \Big\} \nonumber \\
 &\quad + \frac{1}{4} a^2 \delta_V' \Big\{ 
  2D^{\pi S}_{X \eta \eta', X}(Z_V) l(Z_V)
  \nonumber \\
 &\qquad + 2D^{\pi S}_{Y \eta \eta', Y}(Z_V) l(Z_V)
  \nonumber \\
 &\qquad - \Theta^{VF} R^{\pi S}_{XY \eta \eta'}(Z_V) l(Z_V) 
  \Big\} 
  + (V \rightarrow A)
 \Bigg] \nonumber \\
 & + \frac{1}{6} \Big\{ 
    R^{\pi S}_{X \eta}(X_I) \tilde{l}(X_I) 
  + R^{\pi S}_{Y \eta}(Y_I) \tilde{l}(Y_I)
  \Big\} \nonumber \\
 & + \frac{1}{2} a^2 \delta_V' \Big\{
    R^{\pi S}_{X \eta \eta'}(X_V) \tilde{l}(X_V)
  + R^{\pi S}_{Y \eta \eta'}(Y_V) \tilde{l}(Y_V)
  \Big\} \nonumber \\
 & + (V \rightarrow A).
\end{align}
For $m_x = m_y$, we find
\begin{align}
 \label{eq:delta_f_xeqy_pq_21}
 \delta f
 &_{P_F^+, m_x = m_y}^{\mathrm{disc}} = 
 \frac{1}{4} a^2 \delta_V (4-\Theta^{VF}) \Bigg[
   R^{\pi S}_{X \eta \eta'}(X_V) \tilde{l}(X_V)
  \nonumber \\
 &\qquad \qquad \qquad \qquad 
  + \sum_Z D^{\pi S}_{X \eta \eta', X}(Z_V) l(Z_V)
 \Bigg] \nonumber \\
 &+(V \rightarrow A).
\end{align}
%

\subsubsection{\label{subsubsec:su3_q}Quenched case}
In the quenched case, there is no connected contribution, 
Eq.~\eqref{eq:delta_f_con}.
As explained in Refs.\cite{Bernard:1992mk,Bernard:2001yj,Aubin:2003mg},
quenching the sea quarks in the disconnected part can be done by
replacing the disconnected propagator with
\begin{equation}
 D^{a,\mathrm{quench}}_{il}
  =-\frac{\delta_a^\mathrm{quench}}{(q^2+I_a)(q^2+L_a)},\label{Dquench}
\end{equation}
where
\begin{equation}
 \delta_a^\mathrm{quench}=
  \begin{cases}
    4(m_0^2+\alpha q^2)/3 & \text{if $a=I$}\\
    \delta_a & \text{if $a\neq I$.}
  \end{cases}
\end{equation}
Here, note that $I_a$ and $L_a$ represent the squared tree-level masses of
$i\bar{i}$ and $l\bar{l}$ PGBs, respectively, while $I$ represents
the taste-singlet irrep.

Replacing $D^a_{il}$ with the quenched disconnected propagator in
Eq.~\eqref{eq:delta_f_disc} for $m_x \neq m_y$ gives
\begin{align}
 \label{eq:delta_f_xney_quench}
 \delta f
  &_{P_F^+, m_x \neq m_y}^{\mathrm{disc}} \nonumber \\
  & = 
    \frac{\alpha}{6} \Big\{ 
     \frac{Y_I+X_I}{Y_I-X_I}(l(X_I)-l(Y_I)) - X_I \tilde{l}(X_I) - Y_I \tilde{l}(Y_I)
   \Big\} \nonumber \\
  & \qquad + \frac{m_0^2}{6} \Big\{ \tilde{l}(X_I) + \tilde{l}(Y_I) 
     - 2 \frac{l(X_I) - l(Y_I)}{Y_I - X_I} \Big\}
     \nonumber \\
  & + \frac{1}{4}a^2 \delta_V' \Big\{ 2 \tilde{l}(X_V) + 2 \tilde{l}(Y_V)
     - \Theta^{VF} \frac{l(X_V) - l(Y_V)}{Y_V - X_V} \Big\}
     \nonumber \\
  & + (V \rightarrow A),
\end{align}
and for $m_x = m_y$,
\begin{align}
 \label{eq:delta_f_xeqy_quench}
 \delta f
  &_{P_F^+, m_x = m_y}^{\mathrm{disc}} = 
  \frac{1}{4} a^2 \delta_V' (4 - \Theta^{VF}) \tilde{l}(X_V)
  + (V \rightarrow A).
\end{align}

Quenching the sea quarks also affects the analytic terms.
In the quenched version of Eq.~\eqref{eq:delta_f_anal}, there is no 
$L_4$ term of Eq.~\eqref{eq:delta_f_anal}, 
which is coming from the sea quarks.

\subsection{\label{subsec:su2}SU(2) chiral perturbation theory}

We obtain the SU(2) SChPT results from the SU(3) SChPT results using the
prescription of Ref.~\cite{Bailey:2012wb}.  SU(2) chiral perturbation theory
was developed in Ref.~\cite{Gasser:1983yg} and applied to simulation data for
the taste Goldstone decay constants in
Refs.~\cite{Bazavov:2009fk,Bazavov:2009ir,Bazavov:2010yq}.  The results of this
section extend the results of Refs.~\cite{Du:2009ih,Bazavov:2010yq} to the
taste non-Goldstone case.

%
\subsubsection{\label{subsubsec:su2_full}Fully dynamical case}
From Eqs.~\eqref{eq:f_pi_con_su3_full},  
\eqref{eq:f_K_con_su3_full}, and \eqref{eq:f_Kn_con_su3_full}, 
we obtain the connected contributions for the fully dynamical 
1+1+1 flavor case ($m_u \ne m_d \ll m_s$):
\begin{align}
 \label{eq:f_pi_con_su2_full}
 \delta f_{\pi_F^+}^{\textrm{con}} 
  = -\frac{1}{32} \sum_{B} g_B \Big( l(U_B) + 2l(\pi_B^+) + l(D_B) \Big),
\end{align}
\begin{align}
 \label{eq:f_K_con_su2_full}
 \delta f_{K_F^+}^{\textrm{con}} 
  = -\frac{1}{32} \sum_{B} g_B \Big( l(U_B) + l(\pi_B^+) \Big).
\end{align}
\begin{align}
 \label{eq:f_Kn_con_su2_full}
 \delta f_{K_F^0}^{\textrm{con}} 
  &= -\frac{1}{32} \sum_{B} g_B \Big( l(D_B) + l(\pi_B^+) \Big).
\end{align}

For the disconnected contributions, we find from 
Eqs.~\eqref{eq:f_pi_disc_su3_full}, \eqref{eq:f_K_disc_su3_full} and
\eqref{eq:f_Kn_disc_su3_full}:
\begin{align}
 \delta f
  & _{\pi_F^{+}}^{\textrm{disc}} = \frac{1}{2}\big(l(U_I)+l(D_I)\big)-l(\pi_I^0) \nonumber \\
  & + \frac{1}{4} a^2 \delta_V' \sum_X \Big\{ 
      2 R_{U \pi^0 \eta}^D (X_V) l(X_V) \nonumber \\
  & \qquad \qquad \qquad \qquad  + 2 R_{D \pi^0 \eta}^U (X_V) l(X_V) \Big\}
    \nonumber \\
  & \qquad 
    + \frac{1}{4} a^2 \delta_V' \Theta^{VF} \frac{l(\eta_V)-l(\pi_V^0)}{\eta_V-\pi_V^0}  \nonumber \\
  & + (V \rightarrow A),
\end{align}
\begin{align}
 \delta f
  & _{K_F^{+}}^{\textrm{disc}} = 
    \frac{1}{2}l(U_I) - \frac{1}{4}l(\pi_I^0) \nonumber \\
  & + \frac{1}{2} a^2 \delta_V' \sum_X R_{U \pi^0 \eta}^D (X_V) l(X_V)
    + (V \rightarrow A),
\end{align}
and
\begin{align}
 \delta f
  & _{K_F^{0}}^{\textrm{disc}} = 
    \frac{1}{2}l(D_I) - \frac{1}{4}l(\pi_I^0) \nonumber \\
  & + \frac{1}{2} a^2 \delta_V' \sum_X R_{D \pi^0 \eta}^U (X_V) l(X_V)
    + (V \rightarrow A).
\end{align}

The connected contributions in the fully dynamical 2+1 flavor case
($m_u = m_d \ll m_s$) are
\begin{align}
 \delta f_{\pi_F}^{\textrm{con}} 
  &= -\frac{1}{8} \sum_{B} g_B l(\pi_B), \\
 \delta f_{K_F}^{\textrm{con}} 
  &= -\frac{1}{16} \sum_{B} g_B l(\pi_B).
\end{align}
For the disconnected contributions in the fully dynamical 2+1 flavor case, 
we find
\begin{align}
 \delta f_{\pi_F}^{\textrm{disc}} = 
  & \frac{1}{2} (4-\Theta^{VF}) \Big\{ l(\pi_V) - l(\eta_V) \Big\}
    + (V \rightarrow A), \\
 \label{eq:delta_f_su2_full_21}
 \delta f _{K_F}^{\textrm{disc}} =
  & \frac{1}{4} l(\pi_I) + l(\pi_V) - l(\eta_V)
    + (V \rightarrow A).
\end{align}
%

\subsubsection{\label{subsubsec:su2_pq}Partially quenched case}
Considering $x$ and $y$ to be light quarks ($ m_s \gg m_u, m_d, m_x, m_y$),
the connected contributions 
to the decay constants in the partially quenched 1+1+1 flavor case can be 
obtained by dropping terms corresponding to strange sea quark loops
from Eq.~\eqref{eq:delta_f_con_su3_pq}.
Eqs.~\eqref{eq:delta_f_disc_su3_pq_a}, 
\eqref{eq:delta_f_disc_su3_pq_b} and \eqref{eq:delta_f_disc} give
the disconnected contributions:
\begin{align}
 \label{eq:delta_f_xney_su2_pq}
 \delta f
 &_{P_F^+, m_x \neq m_y}^{\mathrm{disc}} = \sum_{Z} \Bigg[
  \frac{1}{4} \Big\{ 
  D^{UD}_{X \pi^0, X}(Z_I) l(Z_I)
  \nonumber \\
 &\qquad + D^{UD}_{Y \pi^0, Y}(Z_I) l(Z_I)
  - 2 R^{UD}_{XY \pi^0}(Z_I) l(Z_I) 
  \Big\} \nonumber \\
 &\quad + \frac{1}{4} a^2 \delta_V' \Big\{ 
  2D^{UD}_{X \pi^0 \eta, X}(Z_V) l(Z_V)
  \nonumber \\
 &\qquad + 2D^{UD}_{Y \pi^0 \eta, Y}(Z_V) l(Z_V)
  \nonumber \\
 &\qquad - \Theta^{VF} R^{UD}_{XY \pi^0 \eta}(Z_V) l(Z_V) 
  \Big\} 
  + (V \rightarrow A)
 \Bigg] \nonumber \\
 & + \frac{1}{4} \Big\{ 
    R^{UD}_{X \pi^0}(X_I) \tilde{l}(X_I) 
  + R^{UD}_{Y \pi^0}(Y_I) \tilde{l}(Y_I)
  \Big\} \nonumber \\
 & + \frac{1}{2} a^2 \delta_V' \Big\{
    R^{UD}_{X \pi^0 \eta}(X_V) \tilde{l}(X_V)
  + R^{UD}_{Y \pi^0 \eta}(Y_V) \tilde{l}(Y_V)
  \Big\} \nonumber \\
 & + (V \rightarrow A),
\end{align}
and
\begin{align}
 \label{eq:delta_f_xeqy_su2_pq}
 \delta f
 &_{P_F^+, m_x = m_y}^{\mathrm{disc}} = 
 \frac{1}{4} a^2 \delta_V' (4-\Theta^{VF}) \Bigg[
   R^{UD}_{X \pi^0 \eta}(X_V) \tilde{l}(X_V)
  \nonumber \\
 &\qquad \qquad \qquad \qquad \qquad 
  + \sum_Z D^{UD}_{X \pi^0 \eta, X}(Z_V) l(Z_V)
 \Bigg] \nonumber \\
 &+(V \rightarrow A).
\end{align}

The connected contributions to the decay constants in the partially
quenched 2+1 flavor case can be obtained by setting $m_u = m_d$ and
decoupling the strange quark in the 1+1+1 flavor case,
Eq.~\eqref{eq:delta_f_con_su3_pq}.  
From Eqs.~\eqref{eq:delta_f_disc_su3_pq_a} and
\eqref{eq:delta_f_disc_su3_pq_b}, we find the disconnected
contributions in the 2+1 flavor case:
\begin{align}
 \label{eq:delta_f_xney_su2_pq_2p1}
 \delta f
 &_{P^+_F, m_x \neq m_y}^{\mathrm{disc}} = \sum_{Z} \Bigg[
  -\frac{1}{2}
  R^{\pi}_{XY}(Z_I) l(Z_I) 
  \nonumber \\
 &\quad + \frac{1}{4} a^2 \delta_V' \Big\{ 
  2D^{\pi}_{X \eta, X}(Z_V) l(Z_V)
  \nonumber \\
 &\qquad + 2D^{\pi}_{Y \eta, Y}(Z_V) l(Z_V)
  \nonumber \\
 &\qquad - \Theta^{VF} R^{\pi}_{XY \eta}(Z_V) l(Z_V) 
  \Big\} 
  + (V \rightarrow A)
 \Bigg] \nonumber \\
 & + \frac{1}{4} \Big\{ 
    l(X_I) + (\pi_I-X_I)\tilde{l}(X_I) \nonumber \\ 
 & \qquad \qquad + l(Y_I) + (\pi_I-Y_I)\tilde{l}(Y_I)
  \Big\} \nonumber \\
 & + \frac{1}{2} a^2 \delta_V' \Big\{
    R^{\pi}_{X \eta}(X_V) \tilde{l}(X_V)
  + R^{\pi}_{Y \eta}(Y_V) \tilde{l}(Y_V)
  \Big\} \nonumber \\
 & + (V \rightarrow A),
\end{align}
and
\begin{align}
 \delta f
 &_{P^+_F, m_x = m_y}^{\mathrm{disc}} = 
 \frac{1}{4} a^2 \delta_V' (4-\Theta^{VF}) \Bigg[
   R^{\pi}_{X \eta}(X_V) \tilde{l}(X_V)
  \nonumber \\
 &\qquad \qquad \qquad \qquad \qquad 
  + \sum_Z D^{\pi}_{X \eta, X}(Z_V) l(Z_V)
 \Bigg] \nonumber \\
 &+(V \rightarrow A).
\end{align}

Considering $x$ to be a light quark and $y$ to be a heavy quark
($m_s, m_y \gg m_u, m_d, m_x$),
the connected contributions to the decay constants can be obtained
by dropping terms from Eq.~\eqref{eq:delta_f_con_su3_pq} corresponding to strange sea quarks
and $y$ valence quarks circulating in loops; {\it i.e.}, only the
$xu$ and $xd$ terms survive in the sum over $Q$.
From Eqs.~\eqref{eq:delta_f_xney_su2_pq}, 
\eqref{eq:delta_f_xney_su2_pq_2p1}, and \eqref{eq:delta_f_disc}
(or alternatively, Eqs.~\eqref{eq:delta_f_disc_su3_pq_a} and \eqref{eq:delta_f_xney_pq_21}),
we find the disconnected contribution for the partially quenched 1+1+1 flavor case,
\begin{align}
 \label{eq:delta_f_xney_su2_pq_y}
 \delta f
 &_{P_F^+}^{\mathrm{disc}} = \frac{1}{4} \sum_{Z} \Bigg[
  D^{UD}_{X \pi^0, X}(Z_I) l(Z_I)
  \nonumber \\
 &\quad +  2 a^2 \delta_V' 
  D^{UD}_{X \pi^0 \eta, X}(Z_V) l(Z_V)
  + (V \rightarrow A)
 \Bigg] \nonumber \\
 & + \frac{1}{4} 
    R^{UD}_{X \pi^0}(X_I) \tilde{l}(X_I) 
  \nonumber \\ 
 & + \frac{1}{2} a^2 \delta_V' 
    R^{UD}_{X \pi^0 \eta}(X_V) \tilde{l}(X_V)
   + (V \rightarrow A).
\end{align}
For the 2+1 flavor case, we find 
\begin{align}
 \label{eq:delta_f_xney_su2_pq_2+1_y}
 \delta f
 &_{P_F^+}^{\mathrm{disc}} = \sum_{Z} \Bigg[
  \frac{1}{2} a^2 \delta_V'
   D^{\pi}_{X \eta, X}(Z_V) l(Z_V)
  + (V \rightarrow A)
 \Bigg] \nonumber \\
 & + \frac{1}{4} \Big\{ 
    l(X_I) + (\pi_I-X_I)\tilde{l}(X_I)
  \Big\} \nonumber \\
 & + \frac{1}{2} a^2 \delta_V' 
    R^{\pi}_{X \eta}(X_V) \tilde{l}(X_V)
  + (V \rightarrow A).
\end{align}

\section{\label{sec:conclusion}Conclusion}
Our results for the decay constants are given compactly by Eq.~\eqref{eq:ftot} with Eqs.~\eqref{eq:delta_f_rooted} through \eqref{eq:delta_f_anal_rooted}; they reduce to those of Ref.~\cite{Aubin:2003uc} in the taste Goldstone sector.  
The only new LECs are those parametrizing the analytic corrections proportional to $a^2$; the SO(4)-violating contributions are independent of those in the masses.  As shown in Table~\ref{tab:coeff}, the factors $\Theta^{BF}$ multiplying the disconnected pieces of the propagators $D_{xy}^{V,A}$ differ from the coefficients in the taste Goldstone case, but no new LECs arise in the loop diagrams.  In SU(2) chiral perturbation theory with a heavy valence quark, the chiral logarithms are the same in all taste channels; only the analytic $\mc{O}(a^2)$ corrections differ.

Results for special cases of interest can be obtained by expanding the disconnected pieces of the propagators in Eq.~\eqref{eq:delta_f_disc}.  For the fully dynamical case with three non-degenerate quarks, the loop corrections in the SU(3) chiral theory are in Eqs.~\eqref{eq:f_pi_con_su3_full}-\eqref{eq:f_Kn_disc_su3_full}.  Results in the isospin limit are in Eqs.~\eqref{eq:f_pi_con_su3_full_21}-\eqref{eq:delta_K_pi_su3_21_b}.  For the partially quenched case with three non-degenerate sea quarks, loop corrections in the SU(3) chiral theory are in Eqs.~\eqref{eq:delta_f_con_su3_pq}-\eqref{eq:delta_f_disc_su3_pq_b}.  Results in the isospin limit are in Eqs.~\eqref{eq:delta_f_xney_pq_21}-\eqref{eq:delta_f_xeqy_pq_21}.  For the quenched case the results are in Eqs.~\eqref{eq:delta_f_xney_quench}-\eqref{eq:delta_f_xeqy_quench}.  Results in SU(2) chiral perturbation theory are in Eqs.~\eqref{eq:f_pi_con_su2_full}-\eqref{eq:delta_f_su2_full_21} and Eqs.~\eqref{eq:delta_f_xney_su2_pq}-\eqref{eq:delta_f_xney_su2_pq_2+1_y}.
These results can be used to improve determinations of the 
decay constants, quark masses, and the Gasser-Leutwyler LECs by analyzing lattice data from taste non-Goldstone channels.

\begin{acknowledgments}
W.~Lee is supported by the Creative Research
Initiatives program (2012-0000241) of the NRF grant funded by the
Korean government (MEST).
W.~Lee acknowledges support from the KISTI supercomputing
center through the strategic support program [No. KSC-2011-G2-06].
\end{acknowledgments}

%
%

%
%
\bibliographystyle{apsrev} 
\bibliography{ref} 
\end{document}